\documentclass[12pt]{iopart}

\usepackage{graphicx}% Include figure files
\usepackage{dcolumn}% Align table columns on decimal pointÊ
\usepackage{bm}% bold math

\usepackage{setspace}
\usepackage{graphicx}

\newcounter{sub}
\newcounter{subeqn}[sub]
\renewcommand{\thesubeqn}{\alph{subeqn}}
\renewcommand{\theequation}{\thesub\thesubeqn}

\def\be{\begin{equation}}
\def\ee{\end{equation}}

\def\lp{\left(}
\def\rp{\right)}

\def\st{\stepcounter{sub}}

\def\bea{\begin{eqnarray}}
\def\eea{\end{eqnarray}}

\newcommand\no{\nonumber}

\begin{document}

\title[Modeling transcription factor binding]{Modeling transcription factor binding events to DNA
using a random walker/jumper
representation on a 1D/2D lattice with different affinity sites }

\author{Vahid Rezania$^{1,2}$, Jack Tuszynski$^1$ and Michael Hendzel$^1$}
%\address[Also at ]{Department of Science, City Center Campus, Grant MacEwan College, Edmonton, AB T5J 2P2, Canada.}
\address{1- Division of Experimental Oncology, Cross Cancer Institute\\
11560 University Avenue, Edmonton, AB T6G 1Z2
Canada \\
2 - Institute for Advanced Studies in Basic Sciences,
          Zanjan 45195, Iran.}
\ead{vrezania@phys.ualberta.ca}
%\homepage{http://fermi.phys.ualberta.ca/~vrezania}

% Revision date - uncomment to exclude date in the final version

% Running head
%\pagestyle{myheadings}
%\markright{Transcription factors walker/jumper}

\begin{abstract}
Surviving in a diverse environment requires corresponding organism responses.
At the cellular level, such adjustment relies on the transcription factors (TFs) which
must rapidly find their target sequences amidst a vast amount of non-relevant
sequences on DNA molecules. Whether these transcription factors locate their target
sites through a 1D or 3D pathway is still a matter of speculation. It has been suggested that the optimum search
time is when the protein equally shares its search time between 1D and 3D diffusions.
In this paper, we study the above problem using a Monte Carlo simulation by considering a very
simple physical model. A 1D strip, representing a DNA, with a number of low affinity
sites, corresponding to non-target sites, and high affinity sites, corresponding to
target sites, is considered and later extended to a 2D strip. We study the 1D and
3D exploration pathways, and combinations of the two modes by considering three different
types of molecules: a walker that randomly walks along the strip with no dissociation;
a jumper that represents dissociation and then re-association of a TF with the strip at later time at a distant site;
and a hopper that is similar to the jumper but it dissociates and then re-associates at a faster rate than the
jumper.   We analyze the final probability distribution of molecules for each case and find that
TFs can locate their targets fast enough even if they spend 15\% of their search time
diffusing freely in the solution.   This indeed agrees with recent experimental results obtained by Elf et al. \cite{Elf07} and
is in contrast with theoretical expectation.

\end{abstract}

%\pacs{}
\maketitle

%\emph{Key words:} Transcription factor;  DNA;  chromatin;  binding affinity;
 %transcription regulation; random walk}

\section{Introduction}
\noindent
The development of an adult animal from a single cell relies upon the ability of
sequence-targeted DNA regulatory proteins
to coordinate the expression of genes in a development- and tissue-specific manner.
These sequence-specific DNA binding
transcription factors (TFs) must locate target sequences amidst a vast amount of
non-relevant sequences.    Surprisingly, the
binding processes to specific sites happen at very fast rates \cite{Rig70},
approximately 100-1000 times faster
than the upper limit of a diffusion-controlled motion of molecules with the same
size \cite{BV85,Shi99}.
The mechanism(s) whereby these regulatory proteins find their target sites in
long DNA molecules in such a rapid way has been the
subject of extensive theoretical and experimental investigations.  Yet it is
currently still a matter of speculation.  See review by Halford and Szczelkun \cite{HS02}.

In one scheme known as a `sliding' or `scanning' mechanism, the DNA binding
protein binds randomly at any site on the DNA and then
translocates along the sequence until it finds its target \cite{VB89,Shi99}.
In this scenario proteins
move along the DNA by 1D diffusion-controlled motion without losing their
contacts with the DNA.   Each time the protein may
move forward or backward (with equal probability) by taking one step along the DNA.
This is equivalent to the symmetrical random
walk problem in 1D.

The second mechanism, however, involves random walks between disassociation and
re-association events in solution.  In this mechanism two different modes of behavior are
available to proteins, namely the so-called  `hopping'  and  `jumping'
mechanisms, whereby proteins dissociate
from the DNA and move through the solution by 3D
diffusion-controlled motion before they can re-associate with another site on the
same DNA \cite{Ber81,Sta00}.
Hopping refers to the case when the re-association occurs at
a distance, say, $\leq 20$ bp while jumping implies re-association at $> 20$ bp from the
dissociated site.    Although hopping and jumping rates depends on the DNA topology
(linear, folded, etc.), the overall
rate of hopping would be higher than that of jumping \cite{HS02}.
This is due to the low diffusivity
of these macromolecules in solution (diffusion constant
$\leq 10$ $\mu$m$^2$ s$^{-1}$), so that the majority of
re-associations would happen at or very near to the dissociated site \cite{VB89}.

For those proteins with two or more binding sites
(such as the Lac repressor, the \emph{Sfi}I and
\emph{Ngo}MIV endonucleases) on a
single DNA molecule, `intersegmental transfer' is considered as the
third way that proteins move from one site to another.
In this mechanism, a bound protein transiently binds to another site at
the same time.   After releasing, the protein may
either remain at its initial binding site or move to a new site \cite{Mil01}.
This mechanism, however, requires
the juxtaposition of two sites within an interval shorter than the protein-DNA
reaction radius in 3D space.    It is shown that
the intersegmental transfer mechanism is more likely for those proteins with two
binding sites that are separate form each
other by $> 300$ bp \cite{Rin99,PV00}.

The sliding mechanism has been more attractive because of the assumption that
diffusion along the length of the DNA is more
rapid than a 3D search \cite{Rig70,RE74,VB89,Shi99}.  Several
experimental strategies have been developed to determine whether or not the facilitated
diffusion in 1D is the dominant mechanism.
In one procedure, a group of DNA molecules of different lengths that each has a
copy of the target-site is considered
\cite{Jac82,Ehb85,Jel96,JP98,Shi99}.
The results, however, did not rule out 3D diffusion (eg. hopping or jumping) and
just showed that the protein binding
association rate decreases as the DNA molecule gets shorter.  This can be explained
by noting that the binding protein
has less chance to encounter a shorter DNA fragment \cite{HS02}.

In an alternative attempt, Terry et al \cite{Ter85} studied the binding of
the \emph{Eco}RI restriction enzyme
on both linear and circular 388 bp
DNA molecules (fixed length molecules) with two target-sites 51 bp apart.
Interestingly, the protein binding strength depends
on whether the two target sites located closer to one end of DNA molecules or placed
equidistant from both ends.
In the former case, the enzyme showed more frequent cleavages at the innermost target
site while in the latter case no
preference between the target sites was observed.    Binding to the circular DNA
also showed no preference between the
target sites, however, the degree of processivity was approximately two times
higher than that on the linear DNA.
Processivity is referred to as a fraction of total reactions that are cleaved at
both target-sites during a single
binding event.   Strikingly, the processivity results show that translocation
through the hopping (3D) mechanism must be
more efficient than the sliding (1D) mechanism, otherwise the degree of processivity
would be marginally above that on
the linear DNA \cite{Sta00}.

Combining the above strategies, Stanford and Szcselkun \cite{Sta00}
measured the processivity factor
for a group of DNA molecules with the
same length but each consisted of two \emph{Eco}RV target-sites that their
inter-site spacing varied from one DNA to another ranging
from 54 to 764 bp.  Theoretically, by increasing the inter-site spacing, $n$, the
processivity factor decreases with a rate of $n^2$
for 1D diffusion and $n^{1/2}$ for 3D diffusion \cite{DE84}.   Interestingly,
the resulting processivity
factor did not match the 1D diffusion relationship (by several orders of magnitude) and
more or less agreed with the 3D diffusion mechanism.

Halford and Szcselkun \cite{HS02} also reviewed other experimental studies that
were designed to determine
whether a 1D or a 3D search
is preferred by binding proteins and concluded that binding proteins find their
target sites, both \emph{in vitro} and
\emph{in vivo}, primarily through the 3D diffusion (hopping or jumping).  This
is, of course, in contrast to the
perception that 1D diffusion along DNA molecules is the dominant pathway in locating
target sites.

Several theoretical investigations, however, suggested that the optimum search time for a TF to locate
its target site on the DNA strip is when the protein spends equal time sliding along the DNA and diffusing
freely in the solution \cite{Ber81,SM04}.  A more precise calculation was carried out  by Coppey et al. \cite{Cop04}
who proposed a stochastic model for the protein-DNA interaction by considering a series of one dimensional diffusions
of a single protein interrupted by several random jumps from site to site.
They showed that for a DNA with an intermediate length the protein should spend a little bit more time (about 10\% more)
in the 1D pathway in order to minimize the search time.  For a long DNA, however, they recovered the result by  \cite{Ber81,SM04}.

More recently Elf et al. \cite{Elf07} performed an experimental investigation on lac
repressor (LacI) in a living Escherichia coli cell and directly observed specific
binding of the labeled lac repressor.   Interestingly, using single-molecule tracking
technique, they determined 1D and 3D diffusion constants of dimeric LacI-Venus to
be $D_1 = 0.046 \pm 0.01 $ $\mu$m$^2$ s$^{-1}$
(measured \emph{in vitro}) and $D_3 = 3 \pm 0.3$ $\mu$m$^2$ s$^{-1}$ (measured
\emph{in vivo}
and without its DNA biding domain), respectively.   However, using the mean
square displacement plot, they found that the effective diffusion constant of
LacI \emph{in vitro} is $D_{\rm eff} = 0.4 \pm 0.02 $ $\mu$m$^2$ s$^{-1}$ that is
$10$ times bigger than 1D constant.  From this fact they concluded that TFs
spend 87\% of time sliding along DNA.   This is obviously in contrast with
theoretical results found in \cite{Ber81,SM04,Cop04}.  Furthermore, they measured the residence time of TFs on
non-specific sites is about $0.3$ to $5$ ms and the searching time for a single
TF to find a target site is about 65 to 360 sec.   The residence time
corresponds with probability of staying TF in a particular site.

The above result is more or less agrees with a single molecule spectroscopy done by Pant et al. \cite{Pan04,Pan05} 
and Sokolov et al. \cite{Sok05}.   They studied the relation of the protein binding rate 
with the protein concentration, $C$, in the solution.  They argued that the rate of binding
to the specific site is proportional to $C^2$ and  $C$ for pure 1D and 3D diffusion, respectively.   By considering
bacteriophage T4 gene 32 protein, Pant et al. \cite{Pan04,Pan05} found that at low to moderate protein concentration 
the binding rate to the specific site of a single-stranded DNA is more or less proportional to the $C^2$.   
This shows that most of the time the protein bounded to the DNA.

In this study we perform a numerical simulation of the interaction of binding proteins
with DNA molecules.  Our aim is to
elucidate through numerical modeling whether a 1D or a 3D pathway is more favorable and
type of results can be expected from a combination mode of behavior.
Here the 1D pathway refers to
the case that molecules move along the lattice
without disassociation from the lattice while the 3D pathway refers to the case
that molecules disassociate and
then re-associate as they move from site to site.  We consider
a binding protein as a random walker (hopper/jumper) that walks (hops/jumps) along a
one or two-dimensional lattice representing DNA molecules.
Binding to a site depends on the binding energy between the molecule and the site
that is represented here as an affinity to the site.
The smaller the affinity the bound molecule has the greater the chance to translocate to
another site.
In general, each site may have a different affinity from others.  In this study,
however, we assume two different
affinities for simplicity: low affinity (LA) sites and high affinity (HA) sites.
The generalization to
several non-equal affinity sites is straightforward.  The HA sites can be considered
as target (specific) sites that
a binding protein is searching for.   The LA sites are those non-specific sites that
a protein moves along
them to find its target site.    Furthermore, we assume that each site can have up to a
single molecule bound to it at a given time.

It is necessary to note that based on several experimental studies the model considered here, though is very simplistic, 
is a good approximation for prokaryote (and possibly eukaryote) cells  \cite{Elf07,Pan04,Pan05,Sok05} .

\section{Method}
\noindent
To represent the DNA substrate for TF binding, consider a one dimensional lattice
made of $L$ sites (in total) comprised of low affinity (LA) and
high affinity (HA) sites.  Such a system can be achieved, for example,
by pulling DNA with optical tweezers to avoid DNA folding \cite{Pan04,Pan05,Sok05}.
A molecule, representing a TF, may leave an LA site with a
greater probability than an HA site.
As a result, the probability of finding a molecule in an HA site is greater than that
in an LA site.
In terms of the binding energies, these sites can be represented by several
potential wells placed in sequence along the DNA lattice
in which HA sites have deeper potentials \cite{SM04}.   See Fig. 1.
Here, we assume that 10\% of the of total number of accessible sites to TF proteins are
HA sites distributed among
LA sites in an a priori unknown fashion which is reasonable assumption that can be
verified by inspecting actual DNA sequence data. 

The probability of leaving from an LA or HA site can be estimated by observing the residence
time of a binding protein to non-specific or specific sites.   Using the
fluorescent recovery after photobleaching (FRAP) technique,
the mean residence times for several chromatin-binding proteins have been studied
\cite{ Lev00,Mis00,Pha04}.  The reader is referred to the review by
van Holde and Zlatanova \cite{Van06} for further references.
Phair et al. \cite{Pha04} studied over 20 chromatin proteins and distinguished
two slow and fast
recovery populations with mean residence times ranging from $\sim$ 3 to 6 sec
for the fast population
and from $\sim$15 to 30 sec for the slow population.    They found Jun and XBP
proteins have the
shortest mean residence time of $\sim 2$ sec and H1$^0$ has the longest one $\sim 3$ min.
See Tables 1-3 in
the paper by Phair et al. \cite{Pha04}.
%in agreement with \cite{Mis00}.
Elf et al. \cite{Elf07} also measured the residence time in non-specific sites
is about $5$ ms.
Here, we assumed that the molecule
has a 20\% chance to leave an HA site
while it has a 67\% chance to leave LA sites.
We assume that the probability of leaving a particular binding site, $P$, is
related to the binding energy, $E$ via an Arrhenius relationship:
$P=P_0 \exp(-\Delta E/ kT)$. Here $k$ is the Boltzmann's constant and $T$ is the
temperature. Hence the choice of the probability values for LA and
HA sites corresponds to an energy difference between then of approximately
$1.2 kT$ which is at the low end of the expected levels. Energy differences of
less than $kT$ would not results in a high- and low-affinity designation.

As we mentioned earlier, a molecule can walk or hop/jump along the lattice.
A walker moves one
step at a time without dissociating from the lattice.  The walker moves left or right
with an equal probability (symmetric walker).
In the case of jumpping, however, one molecule dissociates from a site (LA or HA) and
after a random time interval, the same or another molecule
associates at any available site (chosen randomly) along the lattice.
The hopper, in contrast with jumper, re-associates to the
lattice instantly (if the free site is available) within a $100$-site radius from the dissociated site.
As a result, the hopper does not travel too far from the dissociated site.
For the hopper/jumper there is no left or right preference.
In the Appendix we have given background information and mathematical formulas for
the probability distributions for random walkers and random jumpers. The reason for
our choice of a random reassociation site is due to the typical range of values for
the 3D diffusion constant which is on the order of $1~ \mu$m$^2$/s. Together with the
formula: $<x^2> = 6 Dt$ and taking the value of $t$ on the order of $1$ s gives the rms
distance traveled in the surrounding medium on the order of $2-3~\mu$m which indicates
that the freed TF can find itself in an arbitrary location along the DNA upon rebinding.

%As indicated above, the jumper does a random walk through the solution
%(a 2D lattice around the DNA strip consisting of sites with zero affinity)
%and associates with a site at a
%further distance from the dissociated site.

The simulation starts from an initial state made of $M~ (<L)$ molecules that
are randomly distributed along the lattice.
For the walking case, a random number $r$ is chosen and then compared with
the site's probability,
$P_0 =$ ($P_{\rm HA}$ or $P_{\rm LA}$).   The site's probability is plotted
schematically in Fig. 1.
As seen, $100$ HA sites are clustered in the middle of the lattice with
a total $L=1000$ sites.
If $r > P_0$ and its immediate neighbor is free, the walker will leave,
otherwise it will stay.
The immediate left or right neighbor is also chosen randomly using a separate
uniform random distribution.  Note that here we do not assume that the molecule is
trapped in the HA for good.  This assumption can change the final result drastically.  We will back to this point later.
Similarly, in the case of hopping/jumping, a random number $r$ is chosen and
then compared with the site's probability,
$P_0 =$ ($P_{\rm HA}$ or $P_{\rm LA}$).  In this case only, a pool of molecules is also considered such that
a new molecule from the pool may associate to a free site on the lattice and
a molecule may dissociate from the lattice and enter the pool.
The jumper will leave the site when $r > P_0$ and enter the pool of free TF's,
otherwise it will stay bound to the DNA.
%The molecule in the pool then performs a random walk motion.
After a random time interval,
a destination site is chosen randomly using a
separate uniform random distribution.  If the new site is free, a molecule
from the pool will associate to the DNA.
In the case of hopping, the new destination site is chosen within a $100$-site radius from the disassociated site
with a diminishing probability with a distance.  As a result, the adjacent sites to the the disassociated site are more probable.
The above procedure is repeated $N_{\rm sim}$ times.

To represent the problem more realistically, we also study the dynamics of
TF's on a 2D lattice consisting of several
affinity sites.  In this case, in addition to LA and HA sites, a
zero affinity site representing solution (water) is also considered.
See Fig. {\ref{fig:2D}}.
As seen in Fig. {\ref{fig:2D}}, as well as individual HA sites, there are
some clustered HA sites along x- and
y-direction.   Here x-direction refers to the motion along a
particular DNA molecule and y-direction
refers to the motion along different DNA molecules.   The clustered HA sites
in y-direction represent individual HA sites from different DNA molecules.
This type of clustering, in fact, have been observed experimentally and
it supports our model.

\section{Results}
\noindent
In total, over the course of our simulation each molecule,
a walker or a jumper, will have $N_{\rm step}$ (successful or
unsuccessful) movement events.  Note that each step represents a
unit of time in the simulation.    In general we are interested in the total time (or step) that takes the
system reaches to the equilibrium.  It is clear that
the equilibrium distribution occurs when all the HA sites are filled.
To have a better statistic, the simulation runs for $N_{\rm sim}$ times.
%$N_{\rm sim}=10000$ initial states are chosen randomly.
Here, the final state for walkers (jumpers) is found after
$N_{\rm step}=1000,~ 10000,~ 50000$ and $100000$ steps ($10,~100$ and $1000$),
each with $N_{\rm sim}=1000$.  The results are shown in
Figs. \ref{fig:walker}-\ref{fig:ratio}.

%A typical simulation procedure is presenting in figure 1. The upper panel shows the distribution of HA and LA sites.
%The middle panel represents the initial distribution
%of molecules among the lattice. As we expected, the initial distribution is random and uniform.

In Figs. \ref{fig:walker} and \ref{fig:jumper} we assume that the $100$
HA sites are clustered in the middle.
Figure \ref{fig:walker} demonstrates the final distribution of the $105$ walkers
after $N_{\rm step}=1000$ steps (blue circle), $10000$ steps
(green diamond), $50000$ steps (red asterisk) and $100000$ steps (black star).
The distribution is fairly uniform everywhere
except close to the HA sites.  The distribution decreases near the edges,
suddenly increases at the edges and then decreases
as it goes toward the middle.
Interestingly, not all the HA sites have the highest probability, except those who placed
at the edges.   %This can be understood ???
Figure \ref{fig:jumper} presents the probability of finding a randomly
distributed ensemble of
$105$ hopper/jumpers after $N_{\rm step}=10$ steps (blue circle),
$100$ steps (red asterisk), $1000$ steps (green diamond) in the lattice.
It is clear that the equilibrium distribution is obtained after
$N_{\rm jump}\sim 100$ jumps.
As expected, all the HA sites have the
highest probability which is interestingly constant (more or less) for all curves.
The LA sites, however, are shown to have
more or less similar probabilities for each run with a limiting value $\sim 0.33=1-P_0$.
Interestingly,
the walker shows very different behaviors at the final state in comparison with jumpers.
It is clear that the longer the molecules walk or jump, the
closer the final distribution approximates an equilibrium distribution where it
should be $P_{N_{\rm step}} (x)\simeq P_{\rm equilibrium}(x) = 1 - P_0(x)$ \cite{Van81}.
Comparing Figs. \ref{fig:walker} and \ref{fig:jumper} one can say the
hoppers/jumpers reach the final distribution
in a much shorter time
(equivalently smaller steps) than walkers.  As seen, even after
$N_{\rm walk}=100,000$ steps, the
final distribution of walkers does not
match the equilibrium distribution completely.
More interestingly, the square root relation is also observed between the
required number of jump events
and number of walk events
in order to reach the equilibrium distribution, i.e.
$N_{\rm jump} \sim \sqrt{N_{\rm walk}}$.
This is in agreement with observations \cite{Sta00}.

The difference between the 1D (walking) and 3D (jumping) pathways is also
studied for different LA and HA sites distributions.
Figure \ref{fig:single} demonstrates final distributions of walkers and
jumpers when $50$ single HA sites are distributed among
LA sites.  In this case the none of the HA sites is clustered.
%Figure \ref{fig:jw1} shows the results when $300$ HA sites with two
%different depths are clustered in
%the middle.   In Fig. \ref{fig:jw2} and \ref{fig:jw3} the $100$ HA sites
%are initially distributed in discrete and continuous
%Gaussian distributions among the LA sites, respectively.
Figure \ref{fig:jw4} represents the results for randomly distributed sites with
different depths.   Though in all cases the jumpers reach the equilibrium
distribution in a shorter time scale, for single
HA site distribution (Fig. \ref{fig:single}) the walkers reach the equilibrium
distribution faster in comparison with the cases
when the HA sites are clustered.

The above results can be explained by noting that the pure jumping case is
equivalent with a system with high protein concentration.  So there are enough TFs around the DNA
to bind the target sites immediately.   In contrast, in the case of pure walking, the protein concentration of the solution
is zero.   So TFs will find their target sites after scanning more or less all the lattice.  To be more realistic, we consider a mixed case.
Starting from the case of (100\% walker, 0\%  jumper), we increase the jumper's percentage in the system.  Interestingly, we find that
after having a mixture of (85\% walker,15\% jumper), the equilibrium distribution can be achieved
after $N_{\rm step} \sim 100-1000$ steps.  This agrees with the experimental results found by
Elf et al. \cite{Elf07} and contrasts with the theoretical expectation \cite{Ber81,SM04,Cop04}.
In Fig. \ref{fig:mixed} we demonstrate the result of three different mixtures of random walkers and jumpers, (95\%, 5\%), (90\%, 10\%)
and (85\%, 15\%) after $N_{\rm step} =100$.    As a result, if TFs spend about 15\% of their time diffusing freely in the solution,
they can find their target sites in a very short time.

In the above simulations we assumed that the molecules can escape from the HA sites
(though, with smaller probability) and so they    would not be trapped in the HA sites
for good after entering those sites.  In order to see how this assumption would affect the results,
we run some simulations by assuming molecules will be trapped for good after entering HA sites.   Interestingly, we find that
the equilibrium distribution cannot be achieved in a short time (100-1000 steps) unless
more than 80\% of the molecules are jumpers.

It is generally known that a symmetric random walk distribution approaches the
Gaussian distribution after a long run, i.e $N_{\rm step} \rightarrow \infty$.
This is a consequence of the the central limit theorem.
In general, the required number of steps to achieve a desire level of convergence
to the equilibrium distribution
can be estimated using the Berry-Ess{\'e}en theorem \cite{Fel66} as
\st
\be\label{cri}
N \geq \frac{25}{4} \frac{\langle |x|^3\rangle^2}{\langle x^2 \rangle^3}
\frac{1}{\epsilon^2},
\ee
where $\epsilon$ represents the level of convergence.  It is clear that the
criterion \ref{cri} relies
on the ratio between the second, $\langle x^2\rangle$, and third,
$\langle |x|^3 \rangle$, moments
of the random walk/jump probability distribution.    In the Appendix we
calculate the ratio
${\langle |x|^3\rangle}/{\langle x^2 \rangle}^{3/2}$ for three different cases:
a random walk probability $p_{\rm w}(x)$,
a random jump probability $p_{\rm j}(x)$, and a mixed probability $p_{\rm wj}(x)$, with
results of $1.000$, $1.2990$ and $1.1482$, respectively.    Figure \ref{fig:ratio}
represents
${\langle |x|^3\rangle}/{\langle x^2 \rangle}^{3/2}$ ratio as function
of $N_{\rm step}$ for $105$ walkers,
jumpers, hopper and/or mixed for the initial clustered distribution (see Fig. 1).
The results for all runs are in the range of $1.3\pm .03$ which is
close to the random jump distribution .

As mentioned earlier, in order to make the model more realistic,
we also performed several simulations
with a 2D lattice (Fig. \ref{fig:2D}).  In general, one would expect several DNA
molecules are present in the solution at different locations.  As a result, TF's that
are moving in the bulk have now more chances to bind.  Similarly to 1D lattice,
we consider two different molecules: walker and hopper/jumper.   We assume that
a walker can only walk along the DNA (x-direction) except when it encounters
zero affinity sites.  In those sites, the walker can move in both x- and y- directions.
Hoppers/jumpers can move in both directions with no restriction.  We used similar
strategy for walking and hopping/jumping from one site to another as we employed in
the 1D lattice (see above).  Results are shown in
Figs. \ref{fig:2Dwalk}-\ref{fig:2D_wa_ju_mixed}.

Figure \ref{fig:2Dwalk} shows the probability distribution of $120$
walkers after walking for $100$, $10,000$ and $100,000$ steps. To
compare, we also show the expected final distribution of molecules
$P_{\rm equilibrium}(x,y)\sim 1-P_0(x,y)$ for given probability in Fig.
\ref{fig:2D}.  The probability distribution of $120$ jumpers after
jumping for $100$ steps is shown in Fig. \ref{fig:2Djump}. Comparing
with the expected distribution, it is clear that the jumpers reach
to the equilibrium distribution in a shorter time (smaller steps)
than walkers.

Figure~\ref{fig:2D_wa_ju_mixed} demonstrates the probability
distribution of $120$ mixed walkers and jumpers after
walking/jumping for $1000$ steps.  As seen, by increasing the number
of jumpers, the final distribution approaches to the equilibrium
distribution.  This is again in good agreement with the experimental results obtained
by Elf et al. \cite{Elf07}.

\section{Discussion}
It is well known that very complicated processes such as the coordination
of gene expression, cellular metabolism,
and organ and tissue development rely on a more fundamental process,
the regulation of gene transcription through transcription factor binding.
This process coordinates the expression of genes in a
tissue-specific manner during early stages of development and tissue
specification and is crucial in the development of an adult animal from a single cell.
The fundamental question is how do these transcription factors find their
target sites amongst
a large number of non-target sites.  This is still a matter of speculation.
Specifically, experimental results suggest that
transcription factors must locate target sequences at very fast rates \cite{Rig70},
approximately 100-1000 times faster
than the upper limit of a diffusion-controlled motion of molecules with the same
dimension \cite{BV85,Shi99}.  An enormous amount of effort has been spent
trying to understand
whether transcription factors locate
their targets through a 1D (sliding along the DNA ) or 3D (diffusion in solution) pathway.
Though earlier investigations favored the 1D pathway, recent experimental studies suggest
that the mixture of 1D and 3D pathways is more efficient \cite{Elf07}.

In this paper we address the above question using a Monte Carlo simulation.
We consider a 1D
strip with a number of potential wells at different depths that are clustered on the strip.
The strip represents a DNA molecule and each potential well represents a binding site.
In the simplest case we distinguish two potential depths: low affinity (LA)
sites with shallow potential and
high affinity (HA) sites with deep potential.  See Fig. 1.
The generalization to several potential depths is
straightforward.

The strip is located in a pool of molecules (a 2D lattice with zero affinity sites)
that can associate with or
dissociate from it.
In our simulations, however, we consider three different molecules:
a walker that performs a random walk along
the strip without dissociation from it; a jumper that dissociates from
the strip and then re-associates with the strip at a distant site after a random time interval;
and a hopper that dissociates and then re-associates at a faster rate than the jumper.
As a result, the hopper travels less distance
than the jumper before re-association.    The walker represents a
binding protein that slides along the DNA to find its
target site via the 1D pathway.   The jumper, in contrast, is a
protein that uses the 3D pathway (free diffusion in the solution) to locate
its target site on the DNA.

Several simulations with different molecules and different number of steps
are performed: walkers only with
$N_{\rm step} = 1000,~10,000,~50,000$ and $100,000$;
jumpers/hoppers only with $N_{\rm step} = 10,~100$ and $1000$; and a mixture
of walkers, jumpers and hoppers.
Our results are plotted Figs. \ref{fig:walker}-\ref{fig:ratio}.
As seen in Figs. \ref{fig:walker} and \ref{fig:jumper}, jumpers
reach the equilibrium distribution after order of $N_{\rm jump} \sim 100$
steps while the final distribution
of walkers even after $N_{\rm walk} =100,000$ steps is quite different from
the expected equilibrium distribution.
Interestingly, we find that $N_{\rm jump} \sim \sqrt{N_{\rm walk}}$, which is in good
agreement with observations \cite{Sta00}.
In the case of mixed molecules, we examined different populations of
molecules and found that even if 15\% of
the molecule population are not walkers, the results are very similar to
the jumper case. See Fig. \ref{fig:mixed}
that is shown the result of three different mixtures of random walkers and jumpers, (95\%, 5\%), (90\%, 10\%)
and (85\%, 15\%) after $N_{\rm step} =100$.    Our results show that if the TFs spend just about 15\% of their search time
diffusing in the solution they are able to locate their target site in a short period of time.
This indeed agrees with experimental results obtained
in \cite{Pan04,Pan05,Sok05} and specifically by Elf et al. \cite{Elf07} who found
87\% of the time TFs are diffusing along the DNA. However, our results are in contrast with the
theoretical expectation that suggests the optimum search time happens when
the protein spends equal time sliding along the DNA and diffusing
freely in the solution (a 50-50 mixture) \cite{Ber81,SM04,Cop04}.

As expected, similar results are also obtained when we consider
a 2D lattice rather than a simple 1D lattice (Fig. \ref{fig:2D}).   It is more realistic to assume several DNA
molecules are present in the solution at different locations.  As a result, TF's that
are moving in the bulk have now more chances to bind.    Results are shown in
Figs. \ref{fig:2Dwalk}-\ref{fig:2D_wa_ju_mixed}.
We compared the probability distribution of $120$
walkers after walking for $100$, $10,000$ and $100,000$ steps with that of $120$ jumpers after
jumping for $100$ steps as shown in Figs. \ref{fig:2Dwalk} and \ref{fig:2Djump}.  It is clear that the jumpers reach
to the equilibrium distribution in a shorter time (smaller steps) than walkers.

In recent years, however,  some experimental investigations suggest a new
role for chromatin in
transcription factor regulation.    Chromatin consists of a fundamental
unit called the nucleosome,
a disc-shaped octamer of eight histone
proteins, bound to DNA.     Each histone interacts with other histones and DNA
to form the nucleosome.
A chromatin-medited mechanism may enhance fidelity of transcriptional
regulation and control of
gene expression in the cell.
Initial experiments monitoring histone H1-GFP fusion
during FRAP experiments have shown a rapid exchange of the H1 with sites
(with few minutes residence time)
while being statically fixed on chromatin \cite{Lev00,Mis00}.
A number chromatin-binding proteins
have subsequently been monitored by Phair et al. \cite{Pha04}
\emph{in vivo} using FRAP method.
They found that transient binding is a common
property among chromatin-associated proteins.   They also demonstrated
that all of these chromatin-binding proteins
show rapid mobility.   They concluded that these proteins continuously
scan the genome space for appropriate
binding sites through diffusional hopping (the 3D pathway) between chromatin fibers.
Buck and Leib \cite{BL06} also studied repressor-activator protein 1
in the yeast cells and found that these proteins locate their
binding sites via a newly dynamic target specification mechanism.
Chromatin-mediated regulation of accessibility coordinates genome-wide
distribution of DNA
sequence motifs to target sites by remodeling the genome itself.

%\section*{Acknowledgment}
\ack
This research was supported by grants from the ACB, NSERC, CSA and CHIR.  Support from the
Alberta Foundation is gratefully acknowledged by J. A. T.   We thank Romain Sibi for his
assistance with some of our numerical simulations.

%\clearpage

\setcounter{sub}{0}
\setcounter{subeqn}{0}
\renewcommand{\theequation}{A\thesub\thesubeqn}

\appendix
\section{Calculating ${\langle |x|^3\rangle  }/{\langle  x^2 \rangle  ^{3/2}}$ for different distribution functions}

\subsection{Random walk distribution}

The distribution function for a non-equal step size random walk in
the range $[-\ell, \ell]$ can be written as
\st
\be
p_{\rm w}(x) = a \delta(x+\ell)+ b \delta(x-\ell),
\ee
where $\delta(x)$ is the Dirac delta function and $a+b=1$.
The $n^{\rm th}$ moment can be calculated as
\st
\bea
\langle x^n\rangle   &=& \int_{-\infty}^\infty x^n p_{\rm w}(x) dx,\no\\
  &=&  \int_{-\infty}^\infty x^n \lp a \delta(x+\ell)+ b \delta(x-\ell)\rp dx,\no\\
  &=& [(-1)^n a+b]\ell^n,
\eea
and
\st
\bea
\langle |x|^n\rangle   &=& \int_{-\infty}^\infty |x|^n p_{\rm w}(x) dx,\no\\
  &=&  \int_{-\infty}^\infty |x|^n \lp {\over} a\delta(x+\ell)+b\delta(x-\ell)\rp dx,\no\\
  %&=&  a \int_{-\infty}^0 (-x)^n \delta(x+\ell) dx
  %+ a\int_0^{\infty} x^n \delta(x+\ell) dx\no\\
  %&~~~~+&  b\int_{-\infty}^0 (-x)^n \delta(x-\ell) dx
  %+ b\int_0^{\infty} x^n \delta(x-\ell) dx\no\\
  &=&a \ell^n+b\ell^n=\ell^n,
\eea

Therefore, for the random walk distribution we have
\st
\bea
\langle |x|^3\rangle  =\ell^3,\\
\st
\langle x^2\rangle  =\ell^2.
\eea
As a result,
\st
\be
\frac{\langle |x|^3\rangle  }{\langle x^2\rangle  ^{3/2}} =1.
\ee

\subsection{Random jump distribution}
\noindent
The distribution function for a random jump in the range $[-L/2, L/2]$ can be written as
\st
\be
p_{\rm j}(x) = (1/L)\lp {\over} H(x+L/2)-H(x-L/2)\rp,
\ee
where $H(x)$ is the Heaviside function:
\st
\be
H(x) =\left\{ \begin{array}{ll}
                0 & x< 0\\
                1 & x \geq 0
               \end{array}\right.
\ee
The $n^{\rm th}$ moment can be calculated as
\st
\bea
\langle x^n\rangle   &=& \int_{-\infty}^\infty x^n p_{\rm j}(x) dx,\no\\
  &=&  (1/L)\int_{-\infty}^\infty x^n \lp {\over} H(x+L/2)-H(x-L/2)\rp dx,\no\\
  &=& \frac{[1+(-1)^n](L/2)^n}{2(n+1)},
\eea
and
\st
\bea
\langle |x|^n\rangle   &=& \int_{-\infty}^\infty |x|^n p_{\rm j}(x) dx,\no\\
  &=&  (1/L)\int_{-\infty}^\infty |x|^n \lp {\over} H(x+L/2)-H(x-L/2)\rp dx,\no\\
  &=& \frac{(L/2)^n}{(n+1)}.
\eea
Therefore, for the random jump distribution we have
\st
\bea
\langle |x|^3\rangle  =(1/4)(L/2)^3,\\
\st
\langle x^2\rangle  =(1/3)(L/2)^2.
\eea
As a result,
\st
\be
\frac{\langle |x|^3\rangle  }{\langle x^2\rangle  ^{3/2}} =\frac{3^{3/2}}{4}=1.2990\,.
\ee

\subsection{Random walk-jump distribution}

The distribution function for a mixed random walk-jump in the
range $[-L/2, L/2]$ where $\ell \leq L/2$
can be written as
\st
\bea
p_{\rm wj}(x) & =& \alpha p_{\rm j}(x)+\beta p_{\rm w}(x),\no\\
&=& {\alpha\over L}\lp {\over} H(x+L/2)-H(x-L/2)\rp+
\beta\lp {\over} a\delta(x+\ell)+ b\delta(x-\ell)\rp,\no\\
\eea
where $\alpha+\beta=1$.
The $n^{\rm th}$ moment can be calculated as
\st
\bea
\langle x^n\rangle   &=& \int_{-\infty}^\infty x^n p_{\rm wj}(x) dx,\no\\
  &=& \frac{\alpha [1+(-1)^n](L/2)^n}{2(n+1)} +\beta [(-1)^n a+b]\ell^n,
\eea
and
\st
\bea
\langle |x|^n\rangle   &=& \int_{-\infty}^\infty |x|^n p_{\rm wj}(x) dx,\no\\
  &=& \frac{\alpha (L/2)^n}{(n+1)}+\beta\ell^n.
\eea
Therefore, for the random walk-jump distribution we have
\st
\bea
\langle |x|^3\rangle  =(\alpha/4) (L/2)^3+\beta\ell^3,\\
\st
\langle x^2\rangle  =(\alpha/3) (L/2)^2+\beta\ell^2.
\eea
As a result,
\st
\be
\frac{\langle |x|^3\rangle  }{\langle x^2\rangle  ^{3/2}} =\frac{(\alpha/4) (L/2)^3+\beta\ell^3}
{{[(\alpha/3) (L/2)^2+\beta\ell^2]}^{3/2}}.
\ee
For a special case $\alpha=1/2=\beta$ and $\ell=L/2$, we
find ${\langle |x|^3\rangle  }/{\langle x^2\rangle  ^{3/2}}=1.1482$.

\noappendix
% Figure legends
%\clearpage
\section*{Figure Legends}

\begin{center}
\begin{figure}
%\hspace{3cm}
\includegraphics[width=1\hsize]{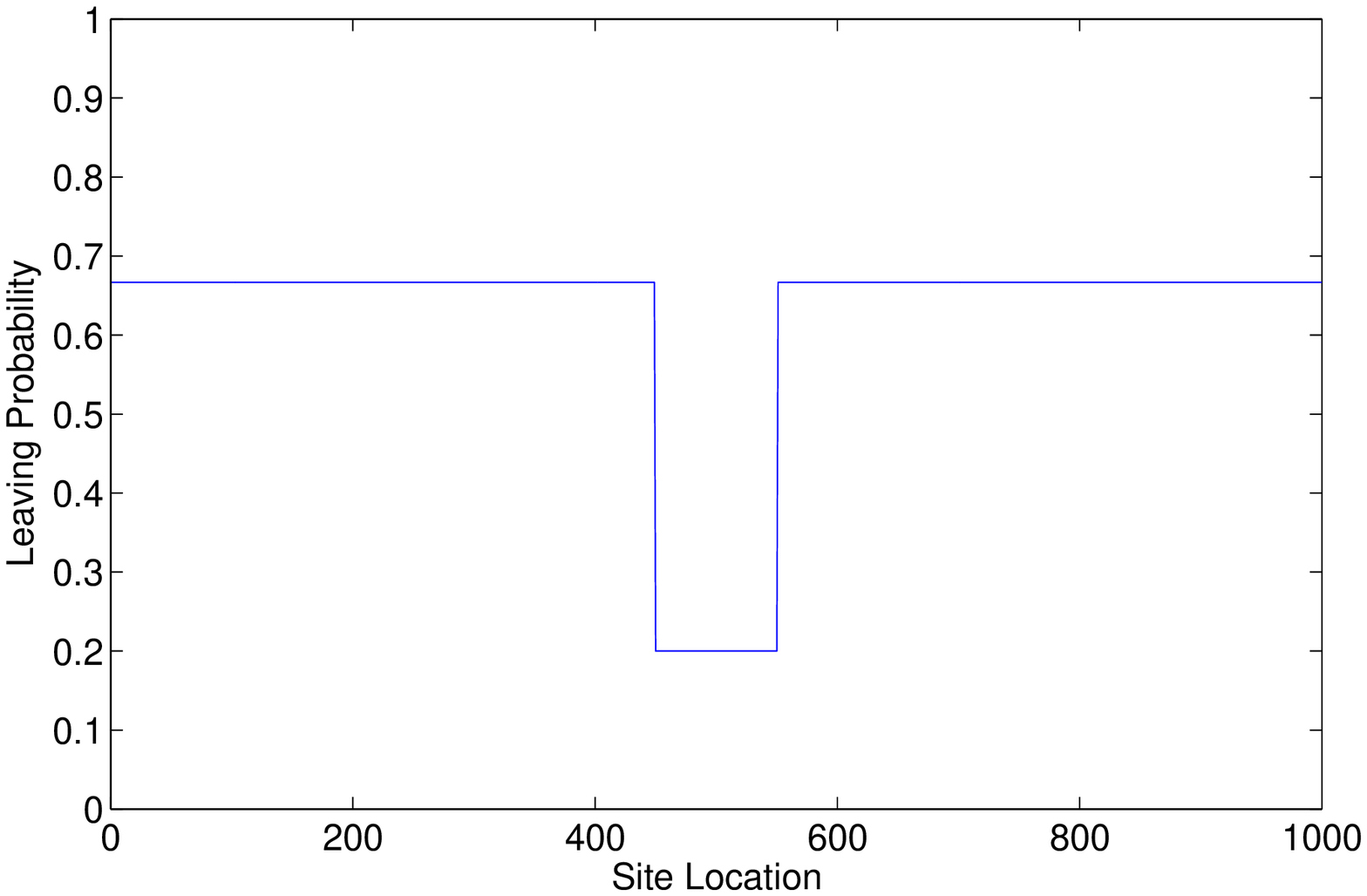}
\caption{   }\label{fig:leaving_prob}
\end{figure}
\end{center}

\subsubsection*{Figure~\ref{fig:leaving_prob}.}
The leaving probability distribution of $L=1000$ sites.  $100$ HA sites are clustered
in the middle.

\subsubsection*{Figure~\ref{fig:2D}.}
The leaving probability distribution $P_0(x,y)$ of $2000$ sites in a
2D lattice. About 70\% of sites are red sites representing the
solution that have zero affinity. The x-direction specifies the
direction along a particular DNA molecule and the y-direction shows
the direction along different DNA molecules.

\subsubsection*{Figure~\ref{fig:walker}.}
Probability distribution of $105$ walkers after walking for
$1000$ steps (circle), $10,000$ steps (diamond), $50,000$ steps (asterisk)
and $100,000$ steps (star).
The HA sites are clustered in the middle.

\subsubsection*{Figure~\ref{fig:jumper}.}
Probability distribution of $105$ jumpers after jumping for
$10$ steps (circle), $100$ steps (asterisk), and $1000$ steps (diamond).
The HA sites are clustered in the middle.

\subsubsection*{Figure~\ref{fig:single}.}
(a) Initial distribution of the $50$ single HA sites (not clustered) with different depths.
(b) Probability distribution of $105$ jumpers after jumping for
$100$ steps for the initial distribution in (a).
Probability distribution of $105$ walkers after walking for (c)
$100$ steps and (d) $10000$ steps for the distribution in (a).

\subsubsection*{Figure~\ref{fig:jw4}.}
(a) Initial randomly distributed binding sites with non-equal depth.
(b) Probability distribution of $105$ jumpers after jumping for
$100$ steps for the initial distribution in (a).
(c) Initial randomly distributed binding sites with non-equal depth.
(d) Probability distribution of $105$ walkers after walking for
$10000$ steps for the initial distribution in (c).

\subsubsection*{Figure~\ref{fig:mixed}.}
Probability distribution of $105$ mixed walkers and jumpers after $100$ steps.
Circle: ($95$\% walker, $5$\% jumper), diamond: ($90$\% walker, $10$\% jumper),
and asterisk: ($85$\% walker, $15$\% jumper).
The HA sites are clustered in the middle.

\subsubsection*{Figure~\ref{fig:ratio}.}
$\langle |x|^3\rangle  /\langle x^2\rangle  ^{3/2}$ value as function of
number of steps for $105$
walkers, jumpers, hoppers and/or mixed molecules.  The $100$ HA sites are
clustered in the middle.
Blue line-circle: walker with $N_{\rm step} = 5000$;
red line-circle: walker with $N_{\rm step} = 10,000$;
black line-circle: walker with $N_{\rm step} = 50,000$;
green line-circle: walker with $N_{\rm step} = 100,000$;
blue line-square: jumper with $N_{\rm step} = 5,000$;
red line-square: jumper with $N_{\rm step} = 10,000$;
blue line-triangle: hopper with $N_{\rm step} = 5000$;
black line-triangle: mixed (50\% walker, 25\% jumper, 25\% hopper)
with $N_{\rm step} = 5000$;
purple line-triangle: mixed (80\% walker, 10\% jumper, 10\% hopper)
with $N_{\rm step} = 5000$.  Those
data with $N_{\rm step}>5,000$ are scaled.

\subsubsection*{Figure~\ref{fig:2Dwalk}.}
(a) Expected final distribution of molecules for given probability
in Fig. \ref{fig:2D}.  Probability distribution of $120$ walkers
after walking for (b) $100$ steps, (c) $10,000$ steps and (d)
$100,000$ steps.

\subsubsection*{Figure~\ref{fig:2Djump}.}
(a) Expected final distribution of molecules for given probability
in Fig. \ref{fig:2D}.  (b) Probability distribution of $120$ jumpers
after jumping for $100$ steps.

\subsubsection*{Figure~\ref{fig:2D_wa_ju_mixed}.}
\begin{center}
\begin{figure}
%\hspace{3cm}
\includegraphics[width=1\hsize]{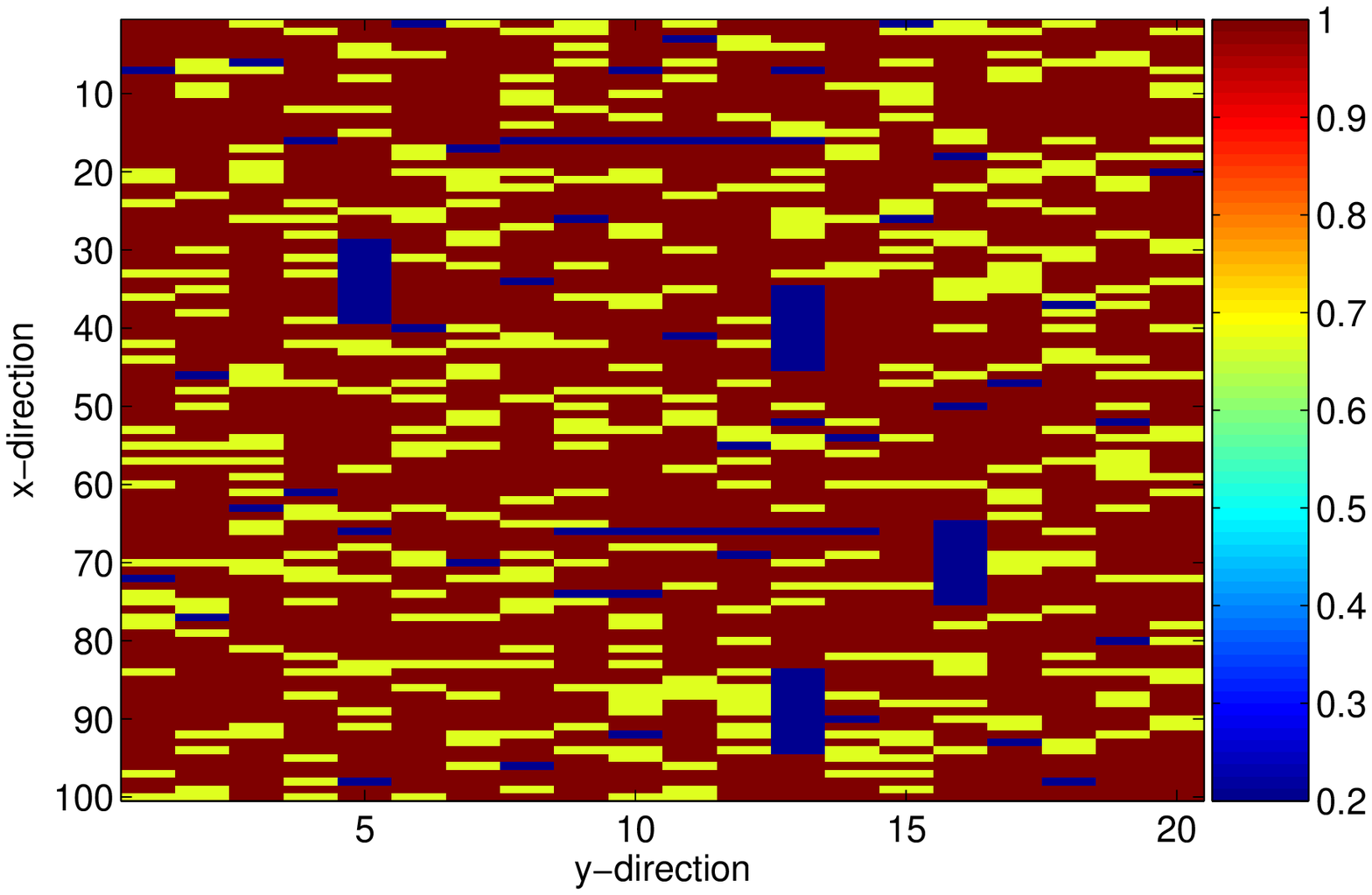}%figure2_2D}
\caption{   }\label{fig:2D}
\end{figure}
\end{center}

\begin{center}
\begin{figure}
%\vspace{-3cm}
\includegraphics[width=1\hsize]{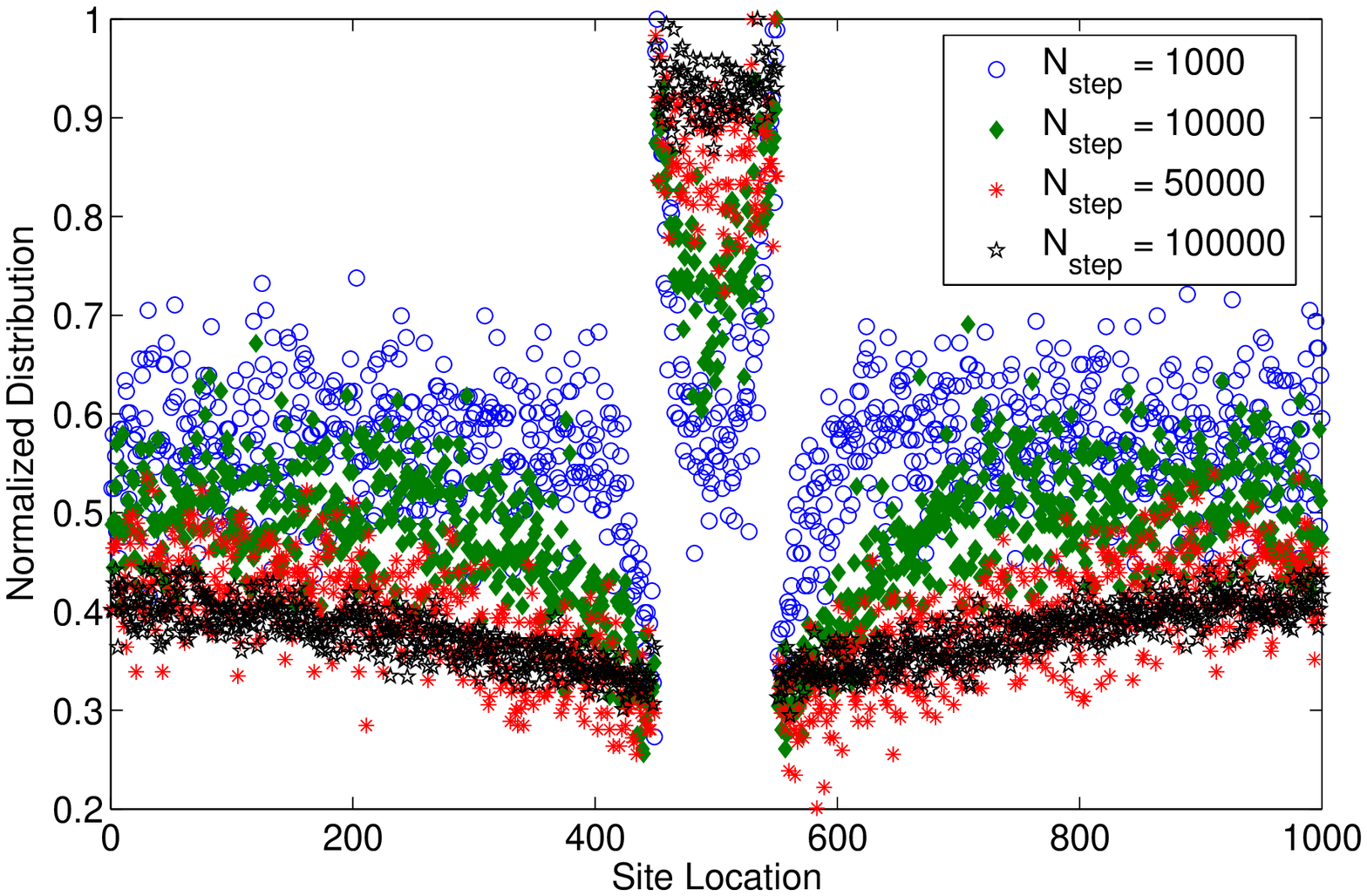}%figure2}
\caption{  }\label{fig:walker}
\end{figure}
\end{center}

\begin{center}
\begin{figure}
%\vspace{-3cm}
\includegraphics[width=1\hsize]{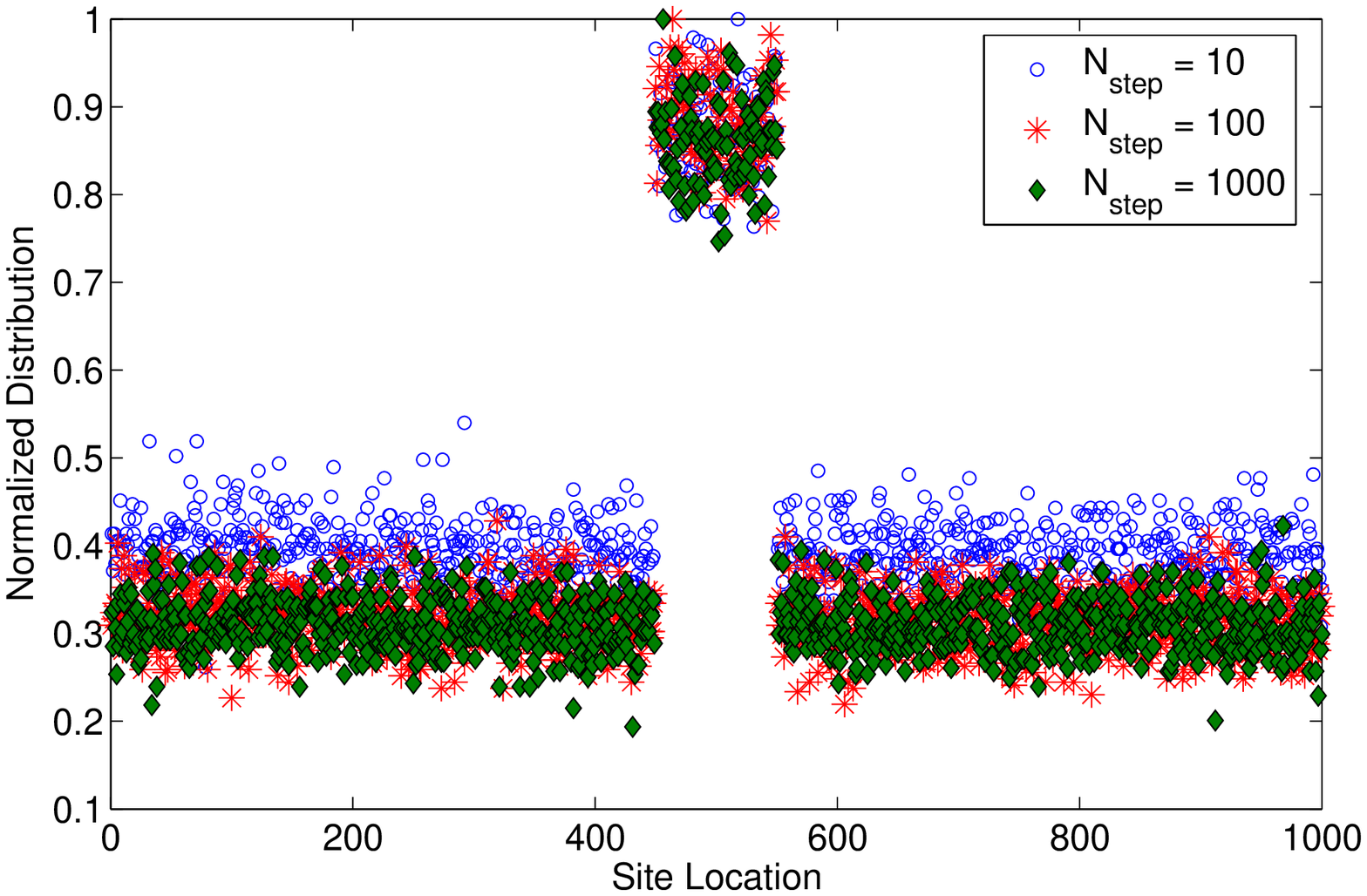}%figure3}
 \caption{} \label{fig:jumper}
\end{figure}
\end{center}

\begin{center}
\begin{figure}
%\vspace{-2.5cm}
\includegraphics[width=1.1\hsize]{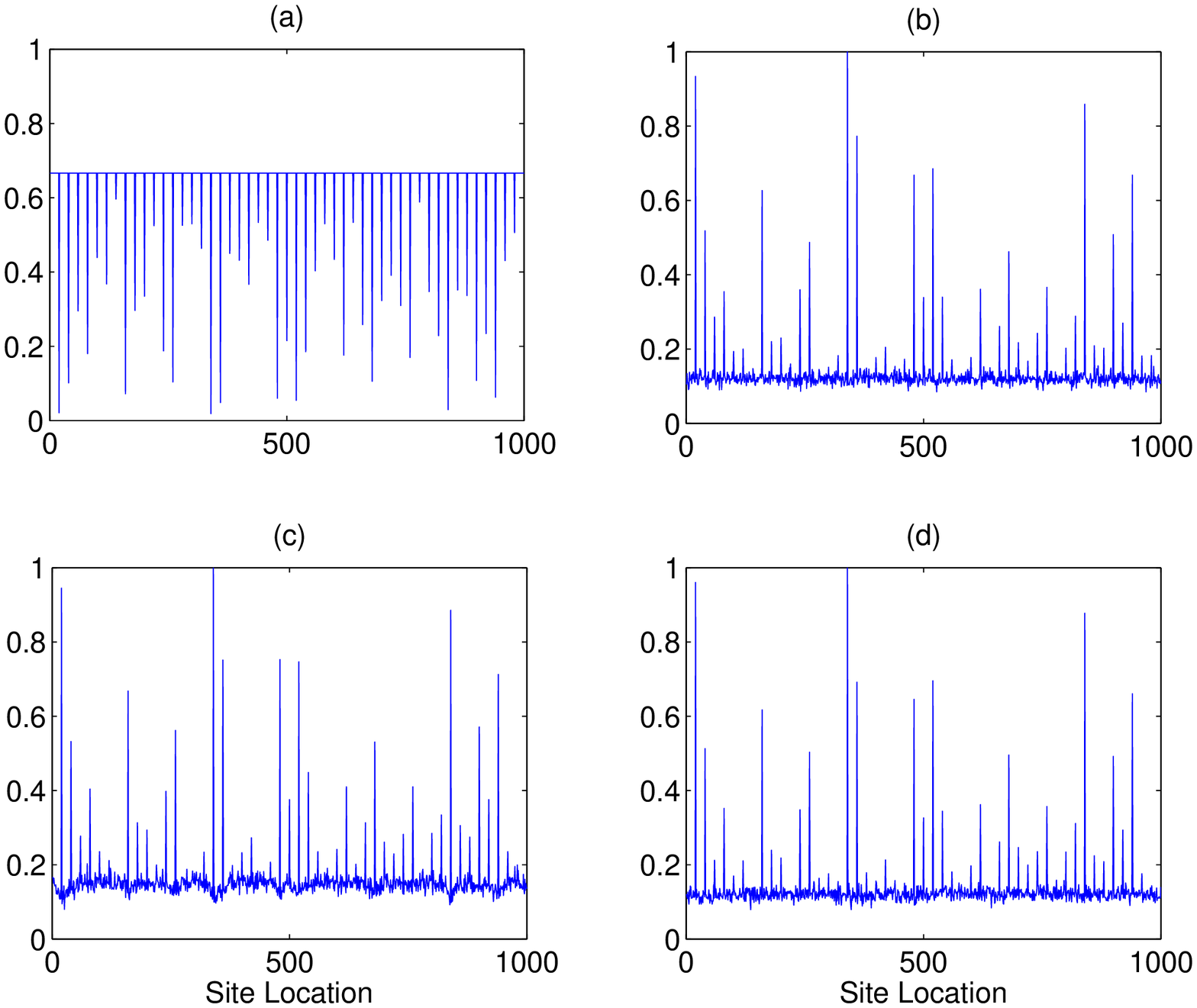}%figure4}
 \caption{} \label{fig:single}
\end{figure}
\end{center}

\begin{center}
\begin{figure}
%\hspace{-2.5cm}
\includegraphics[width=1\hsize]{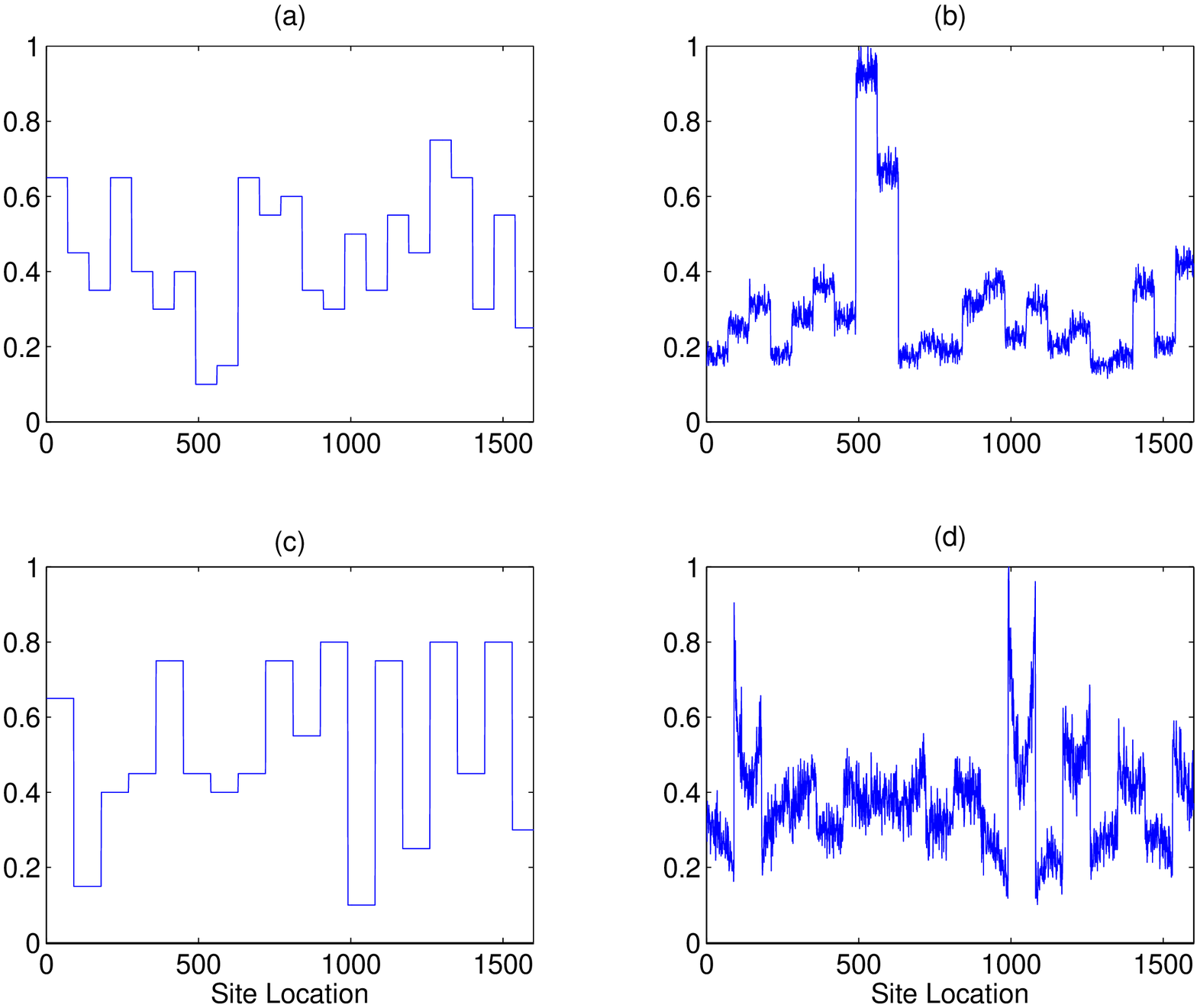}%figure8}
 \caption{} \label{fig:jw4}
\end{figure}
\end{center}

\begin{center}
\begin{figure}
%\vspace{-3cm}
\includegraphics[width=1\hsize]{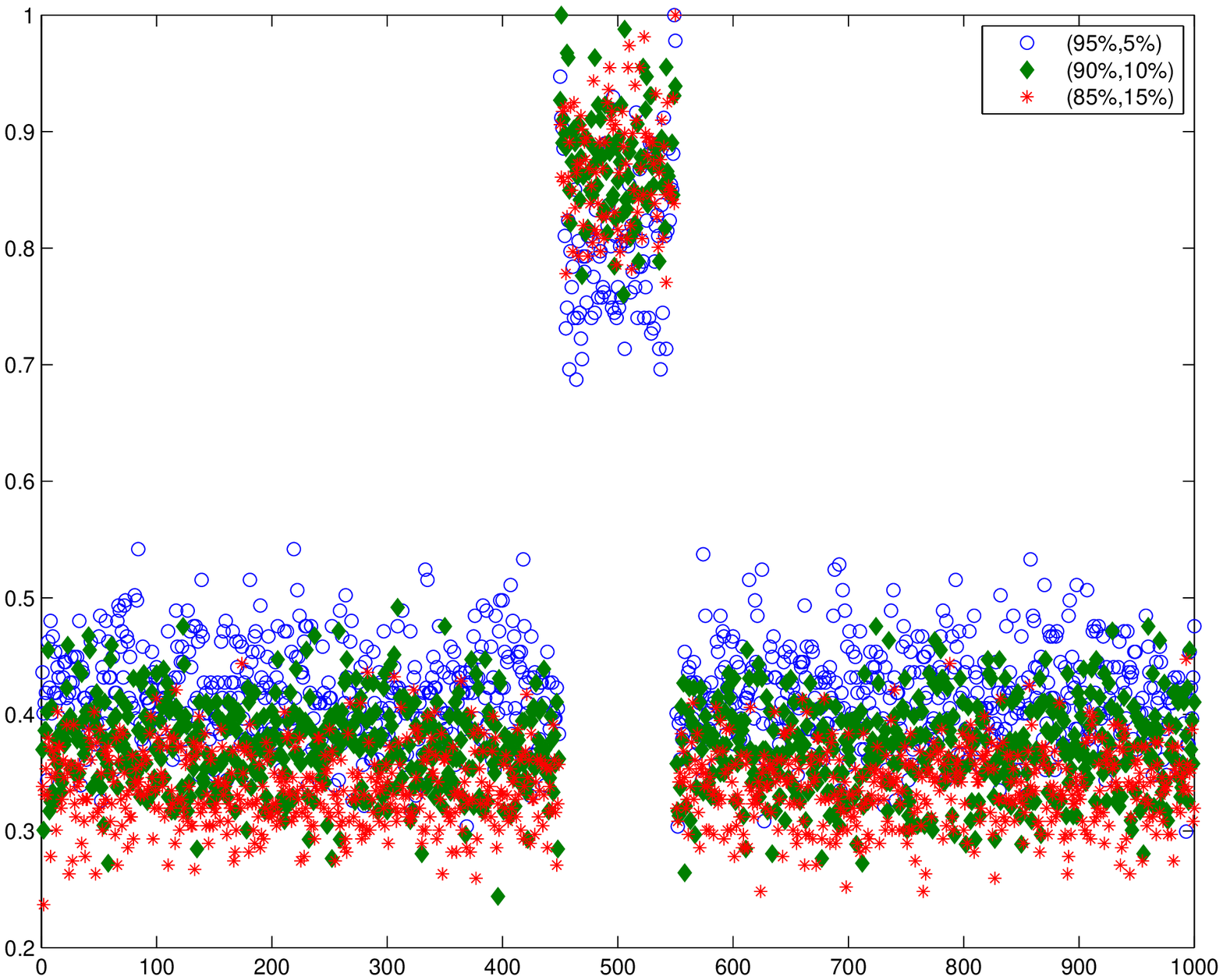}%mixed_case_comparison}
\caption{  }\label{fig:mixed}
\end{figure}
\end{center}

\begin{center}
\begin{figure}
%\vspace{-3cm}
\includegraphics[width=1\hsize]{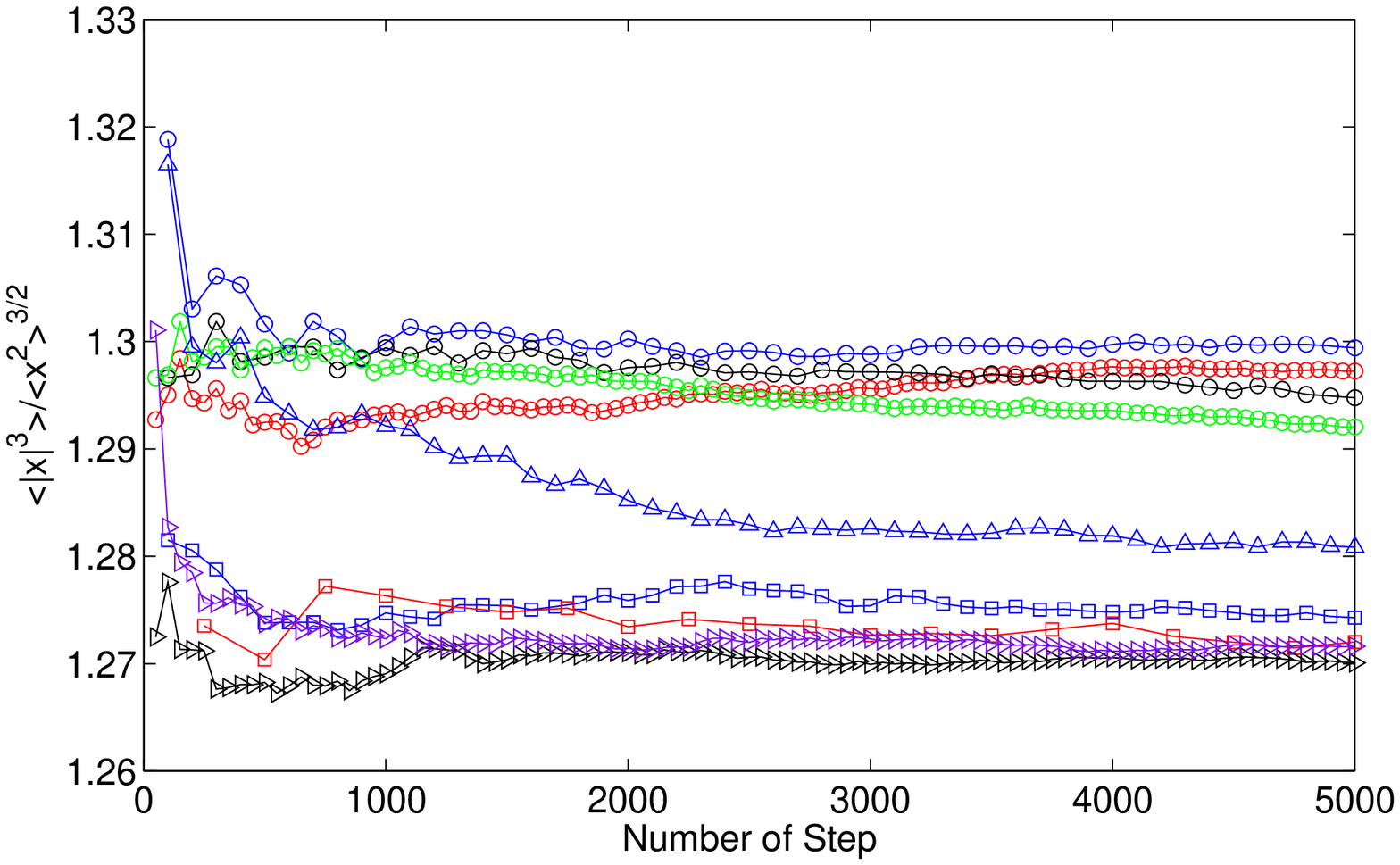}%figure9}
\caption{  }\label{fig:ratio}
\end{figure}
\end{center}

\begin{center}
\begin{figure}
%\vspace{-3cm}
\includegraphics[width=1\hsize]{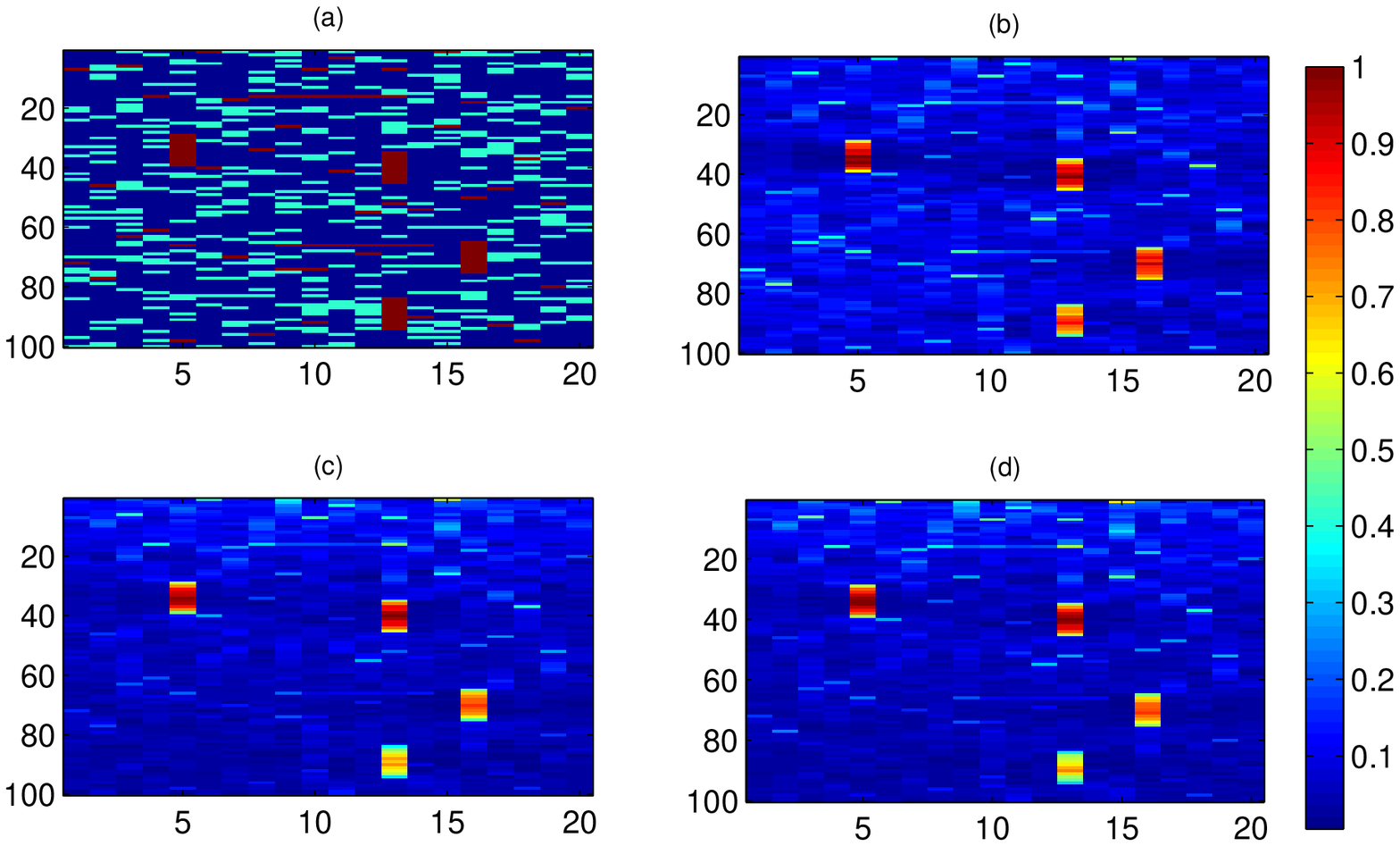}%2Dwalk_120par_100_10000_100000_steps_1000sim}
\caption{  }\label{fig:2Dwalk}
\end{figure}
\end{center}

\begin{center}
\begin{figure}
%\vspace{-3cm}
\includegraphics[width=1\hsize]{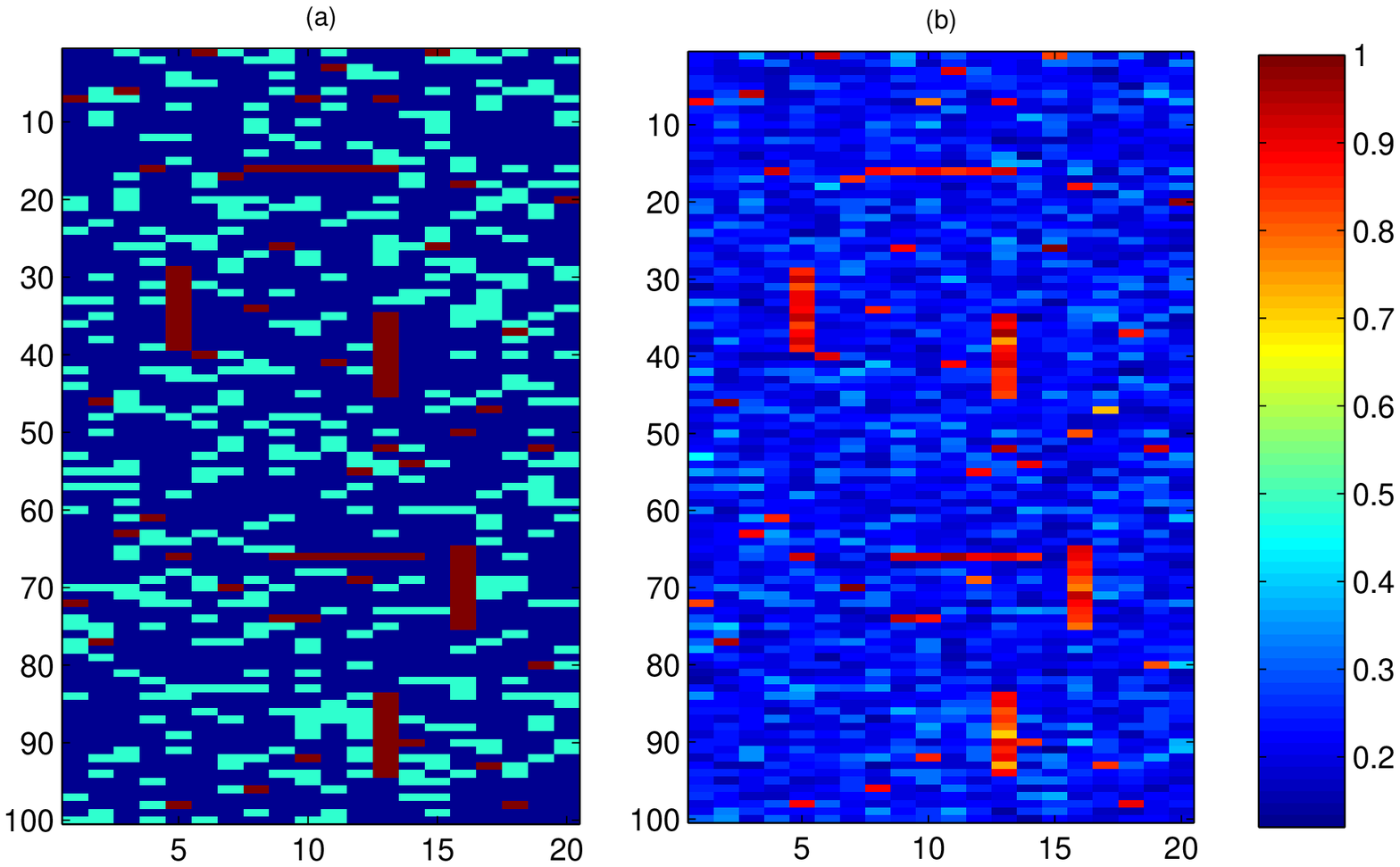}%2Djump_120par_100_steps_1000sim}
\caption{  }\label{fig:2Djump}
\end{figure}
\end{center}

\begin{center}
\begin{figure}
%\vspace{-3cm}
\includegraphics[width=1\hsize]{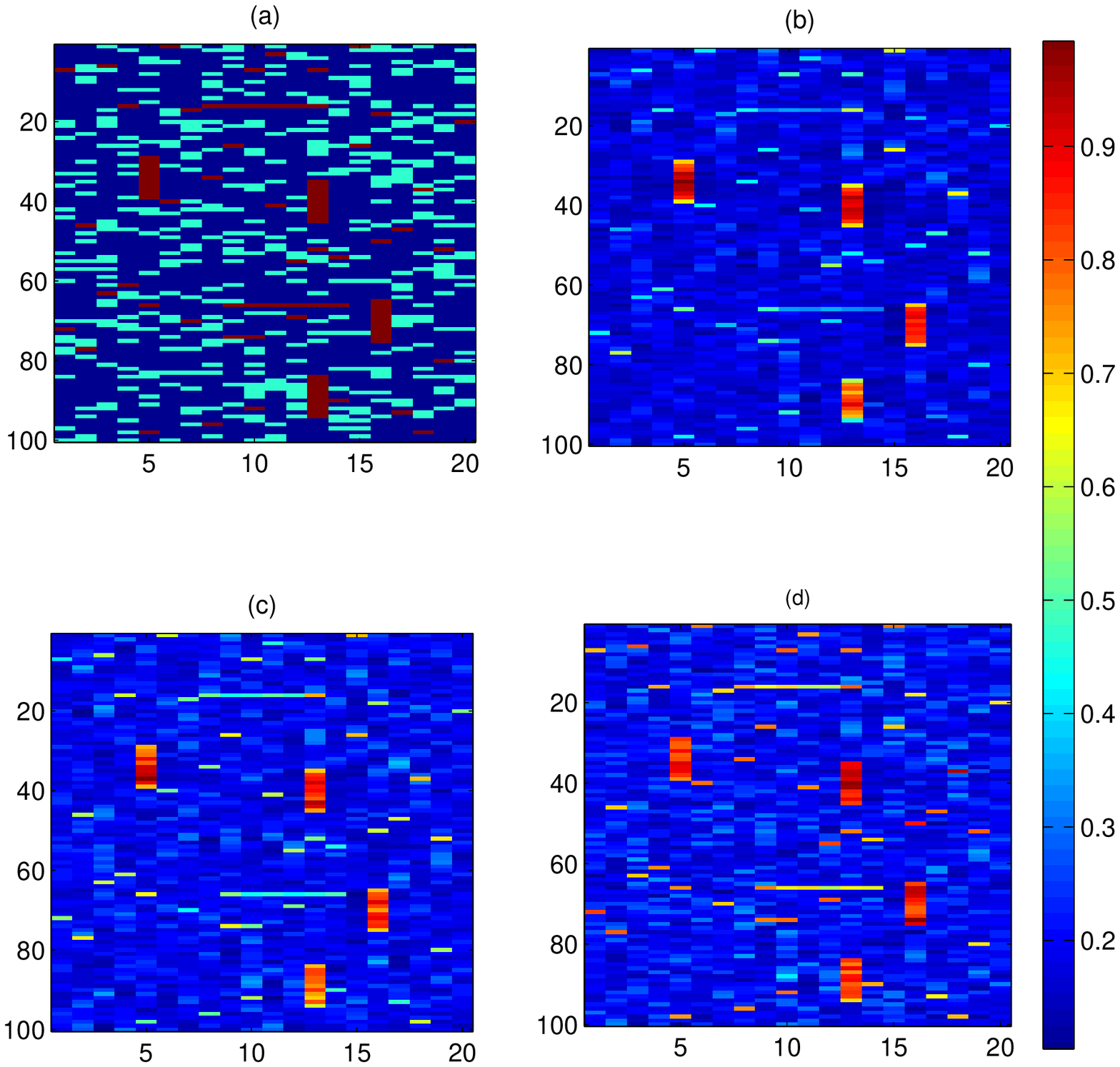}%2Dwa_iu_80_20_50_50_20_80_120par_1000_steps_1000sim}
\caption{  }\label{fig:2D_wa_ju_mixed}
\end{figure}
\end{center}

(a) Expected final distribution of molecules for given probability
in Fig. \ref{fig:2D}.  Probability distribution of $120$ mixed
walkers and jumpers after walking/jumping for $1000$ steps.  (b)
80\% walkers with 20\% jumpers.  (c) 50\% walkers with 50\% jumpers.
(d) 20\% walkers with 80\% jumpers.

\clearpage

\end{document}